\documentclass[a4paper,aps,prl,preprintnumbers,amsmath,amssymb,twocolumn]{revtex4}
\usepackage{bbm}
\usepackage{amsmath}
\usepackage{verbatim}
\usepackage{graphicx}
\usepackage{epsfig}

\newcommand{\tr}{{\rm tr}}

\newcommand{\T}{{T}}

\newcommand{\E}{\mathcal{E}}

\newcommand{\ket}[1]{|#1\rangle}
\newcommand{\bra}[1]{\langle #1|}
\newcommand{\sopt}{s_\infty}

\newcommand{\lmin}{\lambda_{\min}}

\begin{document}
\title{Optimal squeezing and entanglement from noisy Gaussian operations}

\author{Norbert Schuch}
\affiliation{Max-Planck-Institute for Quantum Optics,
 Hans-Kopfermann-Str.\ 1, D-85748 Garching, Germany.}
\author{Michael M.\ Wolf}
\affiliation{Max-Planck-Institute for Quantum Optics,
 Hans-Kopfermann-Str.\ 1, D-85748 Garching, Germany.}
\author{J.\ Ignacio Cirac}
\affiliation{Max-Planck-Institute for Quantum Optics,
 Hans-Kopfermann-Str.\ 1, D-85748 Garching, Germany.}

\begin{abstract}
We investigate the creation of squeezing  via operations subject
to noise and losses and ask for the optimal use of such devices
when supplemented by noiseless passive operations. Both single
and repeated uses of the device are optimized analytically and it
is proven that in the latter case the squeezing converges
exponentially fast to its asymptotic optimum, which we determine
explicitly. For the case of multiple iterations we show that the
optimum can be achieved with fixed intermediate passive
operations. Finally, we relate the results to the generation of
entanglement and derive the maximal two-mode entanglement
achievable within the considered scenario.
\end{abstract}

\date{\today}

\maketitle

Squeezed states are a valuable resource for different fields of
physics. They can increase the resolution of precision
measurements, as exploited in gravitational wave detectors
\cite{Caves}, improve spectroscopic sensitivity
\cite{spectroscopy}, and enhance signal-to-noise ratios
\cite{signal2noise}, e.g., in optical communication. Moreover,
squeezing acts as a basic building block for the generation of
continuous variable entanglement \cite{EPR}, which in turn is a
cornerstone for quantum information purposes. Unfortunately,
squeezing is an expensive resource as well: squeezed states are
hard to create and the involved operations are subject to losses
and noise inevitably restricting the attainable amount of
squeezing. On the other hand passive operations, in quantum
optical setups implemented by beam-splitters and phase shifters,
can often be performed at low cost and they are---compared to the
squeezers---relatively noiseless.

This work is devoted to the question, how can we exploit a given
noisy squeezing device in an optimal way when supplemented by
arbitrary noiseless passive operations. We derive the optimal
strategy for single and repeated use of the squeezing device,
calculate the achievable squeezing and relate it to the maximal
attainable amount of entanglement. To this end we will use a black
box model for the physical squeezing device. This will give us the
possibility to derive optimality results which are equally
applicable to a wide range of physical implementations.

The argumentation will make use of the covariance matrix
formalism, which was mainly developed in the context of continuous
variable states having a Gaussian Wigner distribution---so called
Gaussian states \cite{HolevoBook}. The latter naturally appear in
quantum optical settings (the field of a light mode) as well as in
atomic ensembles (collective spin degrees) and ion traps
(vibrational modes). We restrict ourselves to the natural class of
Gaussian operations, i.e., operations preserving the Gaussian
character of a state \cite{CCR,Geza}. This includes all time
evolutions governed by operators quadratic in bosonic creation and
annihilation operators. All the presented results hold for an
arbitrary number of modes and although it might be reasonable to
think in terms of Gaussian states, we do not have to restrict the
input states to be Gaussian.

\textit{Preliminaries.}---We will begin with introducing the notation
and recalling some basic results \cite{HolevoBook,CCR,Geza,Jens}.
Consider a system of $N$ bosonic modes with respective canonical
operators $ (Q_1,P_1,\dots,Q_N,P_N)=\vec R$. These are related to
the annihilation operators via $a_j=(Q_j+iP_j)/\sqrt{2}$ and
satisfy the canonical commutation relations
$[R_k,R_l]=i\sigma_{kl}\openone$ governed by the symplectic matrix $
\sigma=\bigoplus_{i=1}^N
\mbox{\footnotesize${\left(\begin{array}{cc}0&1\\-1&0\end{array}\right)}$}\;
. $ The displacement $\vec d$ in phase space and the covariance
matrix (CM) $\gamma$ corresponding to a state $\rho$ are then
given by
$$
d_i=\tr[\rho R_i]\mbox{\ \ \ and  \ \
}\gamma_{ij}=\tr\big[\rho\{(R_i-d_i),(R_j-d_j)\}_+\big]\;,
$$
where $\{\cdot,\cdot\}_+$ denotes the anti-commutator.

While for coherent states  $\gamma=\openone$, a state is called
squeezed if its uncertainty in some direction in phase space is
below the uncertainty of the coherent state, i.e., if
$s(\gamma)\equiv\lmin(\gamma)<1$, where $s(\gamma)$ is the
\emph{squeezing} of $\gamma$ measured by its smallest eigenvalue
\cite{mukS}. Note that by this definition, \emph{more} squeezing
means a \emph{smaller} $s$. As the squeezing is independent of the
displacement $\vec d$, we omit it for the remaining part of the paper.

Let us now focus on Gaussian operations. Unitary Gaussian
operations are precisely those realizable by quadratic
Hamiltonians, so that they naturally appear in many physical
systems. In phase space they act as symplectic operations
$S\in\mathrm{Sp}(2N)$ on the covariance matrix $\gamma\mapsto
S^T\gamma S$. Symplectic operations preserve the canonical
commutation relations and are thus given by the group
$\mathrm{Sp}(2N)=\{S\in \mathbb{R}^{2N\times 2N}\,:\,S^\T\sigma
S=\sigma\}$. An important subgroup is given by the group of
orthogonal symplectic transformations $\mathrm K(2N)=\mathrm
O(2N)\cap\mathrm{Sp}(2N)$  \cite{mukS}. Physically, these
correspond to passive operations, which can, in quantum optical
setups, be implemented by beam-splitters and phase shifters
\cite{zeilinger}. Obviously, passive transformations cannot change
the squeezing, since
 elements from $\mathrm K(2N)$ preserve the
spectrum and in particular the smallest eigenvalue of the CM.

We will now introduce the model we use to describe the squeezing
device. In general, a noisy operation $\E$ can be
regarded as a noiseless map on a larger system including the
environment, which is discarded afterwards. If the overall
time-evolution is governed by a quadratic Hamiltonian and the
environment is in a Gaussian state (e.g., a thermal reservoir), it
can be shown that these operations are exactly the ones which act
as
\begin{equation}
\gamma\mapsto\E(\gamma)=X^\T\gamma X+Y\ ,\qquad X^\T i\sigma
X+Y\ge i\sigma \label{eq:channeldef}
\end{equation}
with $X,Y\in\mathbb{R}^{2N\times 2N}$ \cite{CCR}. Here, the
equation on the right hand side ensures complete positivity, i.e.,
guarantees that the operation is physically reasonable. While the
$X^\T\gamma X$ part of $\E$ represents a joint rotation and
distortion of the input $\gamma$, the $Y$ contribution is a noise
term which may consist of quantum
 as well as classical noise. In the following, we will
consider squeezing devices of the type in
Eq.~(\ref{eq:channeldef}), as these are the ones which naturally
appear in many experiments. We will, however, show at the end of
the paper that the results even hold for arbitrary Gaussian
operations (which may include measurements and conditional
operations).

\begin{figure}[tbp]
\begin{center}
\includegraphics[width=7.5cm]{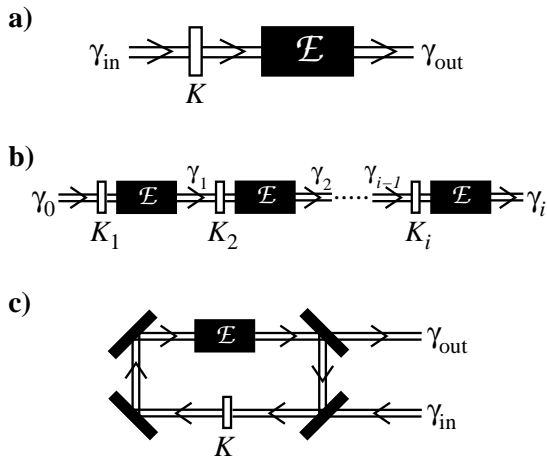}
\end{center}
\caption{Various scenarios for the optimization of squeezing:
\textbf{a)} \emph{single iteration case}: from a given input
$\gamma_\mathrm{in}$, we want to generate as much squeezing as
possible by properly choosing $K$ and using the noisy device $\E$;
\ \textbf{b)} \emph{multiple iteration case}: the device can be
applied repeatedly, and we have to determine the $K_i$'s such as
to maximize the finally obtained squeezing;\ \;\textbf{c)}
\emph{circular setup}: with identical $K_i=K$. }
\label{fig:scenarios}
\end{figure}

The question we are going to investigate  is the following: given
some noisy device $\E$ and the set of passive operations $\mathrm
K(2N)$, how can we generate squeeezing as efficiently as possible
from a given input state? Naturally, this general question can be
asked in several specific ways. First of all, one might ask how
much squeezing can be generated by a \emph{single application} of
the device given a certain initial state, as in
Fig.~\ref{fig:scenarios}a.

The much more interesting question, of course, relates to an
iterative scenario, i.e., how can we generate squeezing as
efficiently as possible by \emph{repeated application} of $\E$
with passive operations $K_i$ inbetween
(Fig.~\ref{fig:scenarios}b)? In this scenario we may either allow
for different $K_i$ or choose them identically as it is for
instance the case in a ring cavity setup
(Fig.~\ref{fig:scenarios}c).

\begin{figure}[tbp]
\begin{center}
\includegraphics[width=7.5cm]{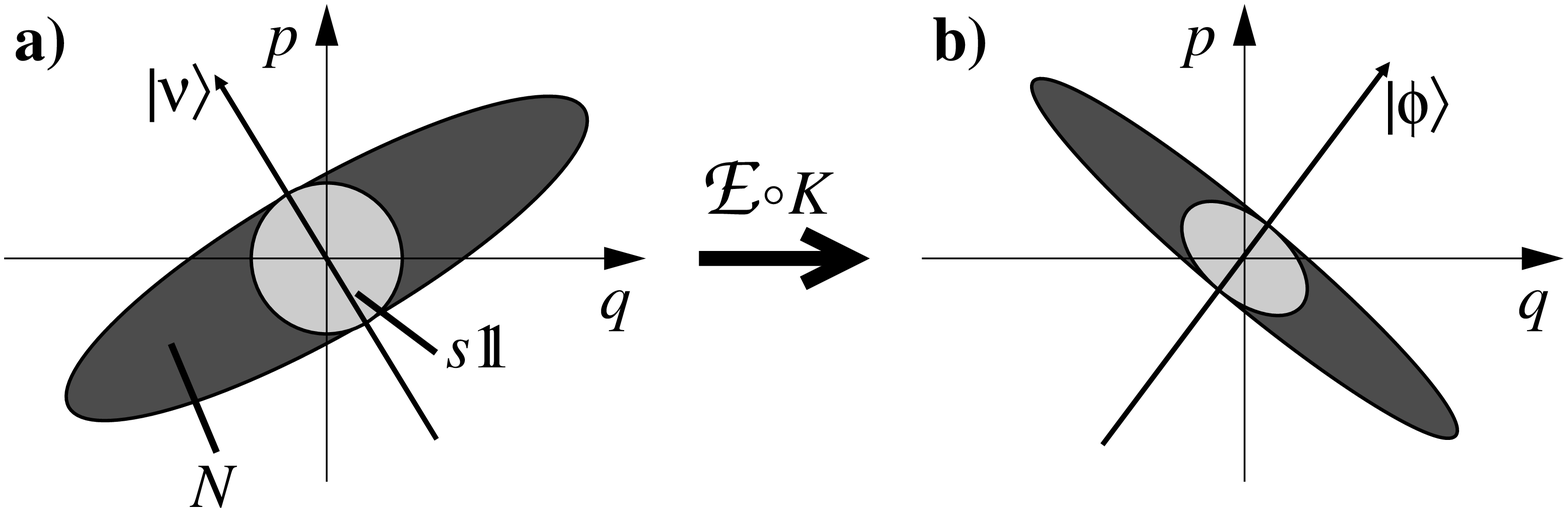}
\end{center}
\caption{Illustration of the single iteration optimality proof.
Left: the input state is decomposed into a coherent kernel
$s\openone$ and into some extra noise $N$ with a null eigenvector
$\ket{\nu}$. Right: after the application of the operation $\E$,
the coherent kernel is squeezed in the direction $\ket{\phi}$. By
choice of $K$, one can achieve $K(X\ket{\phi})\propto\ket{\nu}$,
i.e., the noise $N$ leads to no contribution in the most squeezed
direction. } \label{fig:proofillu}
\end{figure}

\textit{Single iteration.}---In order to prepare for the more
complicated scenarios, let us first have a look at
 the case of a single iteration (Fig.~\ref{fig:scenarios}a), starting
from a given input CM $\gamma$. This is the basic building block
for all the iterative protocols. We now use a formal trick in
order to facilitate the derivation: we split the CM into two parts
(Fig.~\ref{fig:proofillu}a),
\begin{equation}
\gamma=s\openone+N\;, \label{eq:gammadecomp}
\end{equation}
where $s=s(\gamma)$ is the squeezing of $\gamma$. The first part
can be regarded as a ``coherent kernel''  of $\gamma$. It may have
sub-Heisenberg variance (if $s<1$) and it is invariant under
passive operations. The second part is a ``noise term'' $N$ which
ensures that $\gamma$ is a physical state. As $s$ is the smallest
eigenvalue of $\gamma$, $N\ge0$ has a null space.

Let us now see what happens if we rotate $\gamma=s\openone+N$ by
some passive operation $K$ and then send it through $\E$: the
coherent kernel is invariant under $K$, and thus
$$
\gamma'\equiv\E(K^\T\gamma K)=\E(s\openone)+X^\T K^\T NKX\;,
$$
i.e., the action of the squeezing device on the ``coherent
kernel'' plus the ``noise'' part which has been rotated and
squeezed by $X$. Note that the first part no more depends on the
choice of the passive operation; furthermore, as $N\ge0$ and thus
$X^\T K^\T NKX\ge0$, the smallest eigenvalue of $\E(s\openone)$
gives a  bound to the squeezing of the output. In the following,
we show that this  bound can be achieved. Therefore, let
$s_0=s(\E(s\openone))=\lambda_{\min}[sX^\T X+Y]$ be the squeezing
obtained from the input $s\openone$ with corresponding eigenvector
$\ket{\phi}$~\cite{footnote:bra-ket}.
On the one
hand, $s'\equiv s(\gamma')\ge s_0$, and on the other hand,
\begin{eqnarray*}
s'\le\bra{\phi}\E(K^\T\gamma K)\ket{\phi}
    &=&\bra{\phi}\E(s\openone)+X^\T K^\T NKX\ket{\phi}\\
    &=&s_0+\bra{\phi}X^\T K^\T NKX\ket{\phi}\;.
\end{eqnarray*}
Recall that by definition~(\ref{eq:gammadecomp}), $N$ has a null
eigenvector which we denote by $\ket{\nu}$, $N\ket\nu=0$. By
choosing $K$ such that $K(X\ket{\phi})\propto\ket\nu$ (which can
be always done with passive $K$, cf.~\cite{mukS}), the second term
vanishes, and we indeed find that $s'=s_0=s(\E(s\openone))$.

The proof is also illustrated in Fig.~\ref{fig:proofillu}:
Choosing $K$ appropriately ensures that the noise $N$ does not
contribute in the most squeezed direction. Note that for a given
$\E$ it is now straight forward to derive an optimal $K$ and by
exploiting the results of \cite{zeilinger} to decompose it into an
array of beam-splitters and phase shifters.

\begin{figure}[tbp]
\begin{center}
\includegraphics[height=4.3cm]{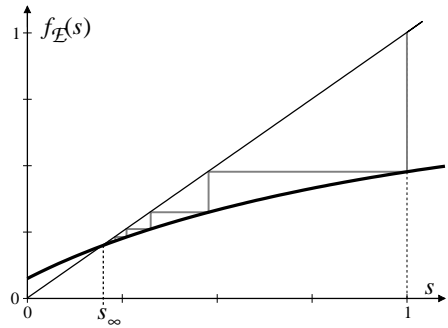}
\end{center}
\caption{The function $s'=f_\E(s)$ describes the optimal output
squeezing $s'$ which can be obtained from input squeezing $s$ by
an operation $\E$. The straight line is the identity, and the gray
zig-zag line describes the evolution of $f$ when using the optimal
strategy for multiple iterations, cf.~Fig.~\ref{fig:scenarios}b. }
\label{fig:sample-f}
\end{figure}

For a single iteration of $\E$ this shows that the optimally
achievable squeezing $s'$ at the output is given by the squeezing
obtained from the non-physical input $s\openone$:
\begin{equation}
s'=f_\E(s):=\lambda_{\min}[sX^\T X+Y]\ . \label{eq:f-def}
\end{equation}
It is highly interesting to note that, therefore, the optimal
value of the final squeezing does \emph{only} depend on the
initial squeezing (and on the properties of $\E$), but on no other
property of the initial state, irrespective of the number of modes
considered. It can be easily checked that the respective function
$f_\E$ in Eq.~(\ref{eq:f-def}) is concave, monotonously
increasing, and $f_\E(0)\ge0$ (Fig.~\ref{fig:sample-f}).

\textit{Multiple iterations.}---The fact that the optimal output
squeezing does only depend on the input squeezing immediately
implies that in the case of multiple iterations the passive
operations $K_i$ can be optimized successively in order to obtain
the global optimum. This is a remarkable result, as in general
problems of this kind require optimization over all parameters
(i.e., over all $K_i$) simultaneously. Graphically, the squeezing
in each iteration step moves along a zig-zag line between
$f_\E(s)$ and the identity, as shown in Fig.~\ref{fig:sample-f}.
The optimal output squeezing $\sopt$ (for
 number of iterations $\rightarrow\infty$) is determined by
$f_\E(\sopt)=\sopt$, $\sopt\ge0$. By inserting the definition
(\ref{eq:f-def}) of $f_\E$ and solving for $\sopt$ we obtain
$$
\sopt=\frac{-1}{\lambda_{\min}[(X^\T X-\openone)Y^{-1}]}\ .
$$
The convergence of $s(\gamma_i)$ to the optimal value is
exponentially fast and bounded from above and below by the slope
of $f_\E$ at $\sopt$ and the slope from $(\sopt,f_\E(\sopt))$ to
the starting point $(s_0,f_\E(s_0))$, respectively. Note that for
$s(\gamma_\mathrm{in})<\sopt$, however,
$s(\gamma_\mathrm{out})>s(\gamma_\mathrm{in})$, so that squeezing
is destroyed by applying the operation $\E$.

In realistic physical scenarios, it might be difficult to tune the
passive operations independently and it is more likely that the
\emph{same} physical device will be passed again and again, e.g.,
in a ring cavity (cf.~Fig.~\ref{fig:scenarios}c), and thus only
one fixed passive operation $K\equiv K_i\ \forall i$ can be
implemented.  In the following, we demonstrate that this is
already sufficient to reach the optimal squeezing $\sopt$. The
proper $K$ is the one which preserves the squeezing at the
optimality point, corresponding to a zig-zag line along the
tangent of $f_\E$ at $\sopt$. The convergence is thus still
exponentially fast.

In order to see how this works, consider the non-physical output
$$
\tilde\gamma=\E(\sopt\openone)=\sopt X^\T X+Y
$$
obtained from the input $\sopt\openone$. By the properties of
$\sopt$, it is clear that $\lambda_{\min}(\tilde\gamma)=\sopt$
with a corresponding \emph{normalized} eigenvector
$\ket{\psi_\infty}$. Now choose $K_\infty$ such that
\begin{equation}
\ket{\psi_\infty}\propto K_\infty X\ket{\psi_\infty}\ .
\label{eq:psi_infty}
\end{equation}
This is exactly the $K$ which preserves the optimality point, as
$\ket{\psi_\infty}$ is the null eigenvector of
$\tilde\gamma-s_\infty\openone$. For any initial state $\gamma$
with $s(\gamma)>s_\infty$, choose the decomposition $
\gamma_0=\sopt\openone+P\; , $ where $P>0$ and
$\lambda_{\min}(P)+\sopt=s_0\equiv s(\gamma_0)$. After one
iteration $\gamma_1=\E(K_\infty^\T\gamma_0 K_\infty)$, we have
\begin{eqnarray*}
s_1&\le&\bra{\psi_\infty}\gamma_1\ket{\psi_\infty}=\sopt+
    \bra{\psi_\infty}X^\T K_\infty^\T PK_\infty X\ket{\psi_\infty}\\
    &\stackrel{(\ref{eq:psi_infty})}{=}&\sopt+\bra{\psi_\infty}P\ket{\psi_\infty}\;
    \underbrace{\bra{\psi_\infty}X^\T X\ket{\psi_\infty}}_{\alpha}\ .
\end{eqnarray*}
As we will show in a moment, $\alpha<1$, and for the squeezing
$s_n\equiv s(\gamma_n)$ after $n$ iterations it holds by recursion
that $ s_n\le\sopt+\alpha^n\bra{\psi_\infty} P\ket{\psi_\infty}$,
which converges exponentially to $\sopt$. From
$$
\sopt\bra{\psi_\infty} X^\T X\ket{\psi_\infty}+
    \bra{\psi_\infty} Y\ket{\psi_\infty}=
    \sopt
$$
it follows that
$$
\alpha\equiv\bra{\psi_\infty} X^\T X\ket{\psi_\infty}=
    \frac{\sopt-\bra{\psi_\infty} Y\ket{\psi_\infty}}
    {\sopt}
$$
which is positive and strictly smaller than one as long as
$\bra{\psi_\infty} Y\ket{\psi_\infty}>0$, which is generically the
case~\cite{footnote:singularY}.

\begin{figure}[t]
\begin{center}
\includegraphics[width=8cm]{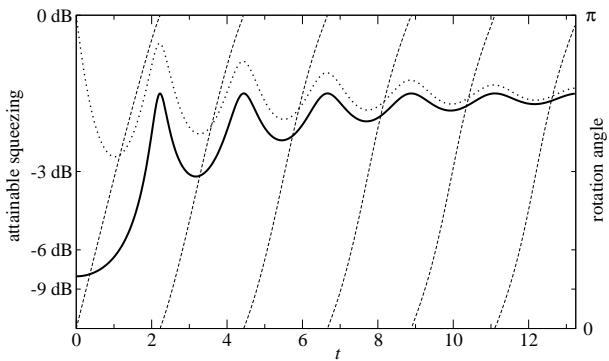}
\end{center}
\caption{Optimization results for a physical device 
$\E_t$ which is given by applying the 
master equation for a time $t$, with 
$H=\frac34a^\dagger a-\frac14a^\dagger a^\dagger+\mathrm{h.c.}$
and $\nu=0.1$ (cf.\ text). The dotted line shows the squeezing obtained
by simple applying $\E_t$ to a coherent state, while the solid line
gives the asymptotically attainable optimum as derived in the paper. The
dashed line shows the rotation angle which leads to the optimal
value.}
\label{fig:example}
\end{figure}

\textit{Example.}---Let us now consider an example which illustrates
how the representation of the operation in
Eq.~(\ref{eq:channeldef}) is related to the master equation
$\dot\rho=i[\rho,\mathcal H]+\mathcal L[\rho]$ of a system. 
From it, one obtains a master equation for the evolution of the
covariance matrix,
\begin{equation}
\label{eq:gammamaster} \dot\gamma=A\gamma+\gamma A^T+N\;.
\end{equation}
For the case of photon losses to a vacuum reservoir, $\mathcal
L[\rho]= \nu(2a\rho a^\dagger-a^\dagger a\rho-\rho a^\dagger a)$,
one obtains $A=2\sigma H-\nu\openone$ and $N=2\nu\openone$, where
$H$ is the Hamiltonian matrix, i.e., $\mathcal H=(Q,P)
H(Q,P)^\dagger$. By integration, one finds that
applying (\ref{eq:gammamaster}) for a time $t$ leads to a map
$\E_t:\gamma\mapsto X^T\gamma X+Y$ with $X=e^{-\nu t}\exp[-2H\sigma t]$
and
$$
Y=X^T\left[2\nu\int_{0}^{t}e^{2\nu\tau}
    e^{-2\sigma H\tau}e^{2H\sigma\tau}\mathrm{d}\tau\right] X\ .
$$
At this point, the results derived in the paper can be applied.
Fig.~\ref{fig:example} shows a one mode example with $\mathcal
H=\frac34a^\dagger a-\frac14a^\dagger a^\dagger+\mathrm{h.c.}$ 
and the noise level $\nu=0.1$.
Note that additional passive operations enhance the attainable squeezing
although the noise $N$ is rotationally invariant.

\textit{Entanglement generation.}---The optimality result on squeezing
generation has direct implications for optimal entanglement generation.
Indeed, it has been shown that the amount of entanglement which can be
generated by passive operations starting from two squeezed Gaussian input
states only depends on their squeezing irrespective of the number of
modes~\cite{WE}; for two modes, e.g., this is done by sending the states
onto a beam splitter after rotating them into orthogonal directions.
Using this result, we can immediately determine how much entanglement we 
can generate from Gaussian inputs using 
a black box $\E$ supplemented by passive operations, namely 
$E_{\mathcal N,\mathrm{opt}}=\log(s_{\text{opt}}^{-1})$, where
$s_{\text{opt}}$ is the maximal squeezing generated by $\E$ and the
entanglement is measured by the logarithmic negativity~\cite{logneg}.
Moreover, by combining the results it is straightforward to explicitly derive the
optimal entangling protocol for any given Gaussian device.

\textit{Acknowledgements.}---We thank Klemens Hammerer and Jens Eisert for valuable
discussions. This work has been supported by the EU
(COVAQIAL), the DFG and the ``Kom\-pe\-tenz\-netz\-werk
Quanteninformationsverarbeitung'' der Bayerischen Staatsregierung.

\textit{Appendix.}---%
In the following, we show that the results obtained 
for channels of the type~(\ref{eq:channeldef}) also hold for the most general type of 
Gaussian channels which may include measurements and postprocessing.
Channels of this type appear, e.g., in the creation of spin squeezing
using quantum nondemolition measurements with feedback~\cite{mabuchi}.
The most general memoryless
operation on $N$ modes can described by a $2N$ mode covariance
matrix $\Gamma$ via the Jamiolkowski isomorphism~\cite{Geza} as
\[
\E(\gamma)=A-C(B+\gamma)^{-1}C^T\,,\mbox{\ where\ }
    \tilde\Gamma=\left(\begin{array}{cc}A&C\\C^\T&B\end{array}\right)\ ,
\]
and $\tilde\Gamma$ is the partial transpose of $\Gamma$. 
Again,
$s\equiv s(\gamma)$, $\ket\psi$ is the eigenvector corresponding
to the smallest eigenvalue of $\E(s\openone)$, and we need to show
that $K$ can be choosen such that $s(\E(s\openone))=s(\E(K\gamma K^T))$,
i.e., that $\bra\psi
C(B+s\openone)^{-1}C^\T\ket\psi=\bra\psi C(B+K\gamma
K^\T)^{-1}C^\T\ket\psi$. This means that for $\ket\chi\equiv
C^\T\ket\psi$, $P\equiv(B+s\openone)>0$, and
$N\equiv\gamma-s\openone\ge0$ with a null eigenspace we have to
find a $K$ such that
\begin{eqnarray}
\label{eq:jam-ch-vanish}
&&\!\!\!\!\!\bra\chi P^{-1}-(P+KNK^\T)^{-1}\ket\chi=\\
&&\!\!\!\!\!\bra\chi P^{-1/2}
    \underbrace{\left[
    \openone-\left[\openone+P^{-1/2}KNK^\T P^{-1/2}\right]^{-1}\right]}_{(*)}
    P^{-1/2}\ket\chi\nonumber
\end{eqnarray}
vanishes. Since $(*)$ has the same null
eigenspace as $P^{-1/2}KNK^\T P^{-1/2}$, the expression (\ref{eq:jam-ch-vanish})
can be indeed made zero by 
choosing $K$
such that $\ket\nu\propto K^\T P^{-1}\ket\chi$ (where
$N\ket\nu=0$).
This proves that for any Gaussian operation the optimal
output squeezing can be computed on
$s\openone$, thus generalizing the results of the paper.


\vspace*{-1em}



\begin{thebibliography}{99}
\vspace*{-1em}
\bibitem{Caves} C.M. Caves, Phys. Rev. D {\bf 23}, 1693 (1981).
\bibitem{spectroscopy} E.S. Polzik, J. Carri, H.J. Kimble, Phys. Rev.
Lett. {\bf 68}, 3020 (1992); N.Ph. Georgiades \emph{et al.},
Phys. Rev. Lett. {\bf 75}, 3426 (1995).
\bibitem{signal2noise} M. Xiao, L.–A. Wu, H.J. Kimble, Phys. Rev. Lett.
{\bf 59}, 278 (1987); 
P. Grangier \emph{et al.}, 
Phys.  Rev. Lett. {\bf 59}, 2153 (1987).
\bibitem{EPR} Z.Y. Ou \emph{et al.},
Phys. Rev.  Lett. {\bf 68}, 3663 (1992); 
C. Silberhorn \emph{et al.},
Phys. Rev. Lett. {\bf 86}, 4267 (2001);
W.  P. Bowen \emph{et al.},
Phys. Rev. Lett. {\bf 90}, 043601 (2003).
\bibitem{HolevoBook} A.S. Holevo, {\it Probabilistic and
statistical aspects of quantum theory}, North-Holland Publishing
Company, 1982.
\bibitem{CCR} B. Demoen, P. Vanheuverzwijn, A. Verbeure, Lett. Math. Phys. {\bf 2}, 161
(1977).
\bibitem{Geza} G. Giedke, J.I. Cirac, Phys. Rev. A {\bf 66}, 032316
(2002); J. Fiurasek, Phys. Rev. Lett. {\bf 89}, 137904 (2002).
\bibitem{Jens} J. Eisert, M.B. Plenio, Int. J. Quant. Inf. {\bf 1}, 479
(2003).
\bibitem{mukS} R. Simon, N. Mukunda, B. Dutta, Phys. Rev. A {\bf 49}, 1567
(1994).
\bibitem{zeilinger} M. Reck \emph{et al.},
Phys. Rev. Lett. {\bf 73}, 58 (1994).
\bibitem{footnote:bra-ket} Note that we use bra-kets to denote
    ordinary vectors. This is \emph{not} a quantum state. 
\bibitem{footnote:singularY} The case of singular $Y$ requires a lengthy
discussion which, however, does not give any new insight, therefore, it has
been omitted.
\bibitem{WE} M.M. Wolf, J. Eisert, M.B. Plenio, Phys. Rev. Lett. {\bf 90}, 047904
(2003).
\bibitem{logneg} G. Vidal, R.F. Werner, Phys. Rev. A {\bf 65}, 032314 (2002).
\bibitem{mabuchi}J.M.\ Geremia, J.K.\ Stockton, and H.\ Mabuchi, Science
\textbf{304}, 270 (2004).
\end{thebibliography}
\end{document}